\documentclass{elsart}
\begin{document}

\begin{frontmatter}
\title{Embedding quantum and random optics in a larger field theory}
\author{Peter Morgan}
\ead{peter.w.morgan@yale.edu}
\address{Physics Department, Yale University, New Haven, CT 06520, USA.}
\ead[url]{http://pantheon.yale.edu/~PWM22}

\begin{abstract}
Introducing creation and annihilation operators for negative frequency components extends the
algebra of smeared local observables of quantum optics to include an associated classical
random field optics.
\end{abstract}

\end{frontmatter}

The free field aspect of quantum optics can be constructed as a quantum field theory using
creation and annihilation operators, $\hat a_g^\dagger$ and $\hat a_f$, that satisfy the
commutation relation
\begin{eqnarray}\label{QCommRel}
  \left[\hat a_f,\hat a_g^\dagger\right]&=&(g,f)
   =-\hbar\int k^\alpha\tilde g^*_{\alpha\beta}(k)\eta^{\beta\nu}
                                          k^\lambda\tilde f_{\lambda\nu}(k)\theta(k_0)2\pi\delta(k_\mu k^\mu)
                                          \frac{\mathrm{d}^4k}{(2\pi)^4},\cr
  \left[\hat a_f,\hat a_g\right]&=&0.
\end{eqnarray}
$f_{\mu\nu}(x)$ and $g_{\mu\nu}(x)$ are complex bivector test functions, which are typically taken to be pure
positive frequency delta functions in quantum optics, even though a distribution is an improper test function
and even though the smeared field operators $\hat\phi_f=\hat a_f+\hat a_{f^*}^\dagger$ are observables only
if $f_{\mu\nu}(x)=f^*_{\mu\nu}(x)$ are real functions.
For each null 4-vector wave-number $k_\lambda$, the expression
$k^\alpha\tilde f^*_{\alpha\beta}(k)\eta^{\beta\nu}k^\lambda\tilde f_{\lambda\nu}(k)$ is negative
semi-definite (taking the metric to be $\eta=\mathrm{diag}[1,-1,-1,-1])$ --- that is,
$k^\lambda\tilde f_{\lambda\nu}(k)$ is a zero or space-like 4-vector --- so $(g,f)$ is an
inner product on the test function space.
A vacuum state $\left|0\right>$ is annihilated by all annihilation operators, $\hat a_f\left|0\right>=0$.
This presentation of quantum optics is a natural consequence of Ref. \cite{MS}, which derives the
inner product between test functions that is given above.

With the above constructions, the smeared field operator $\hat\phi_f$ satisfies microcausality,
$\left[\hat\phi_f,\hat\phi_g\right]=0$, when the supports of the test functions $f$ and $g$ are
space-like separated; for test functions with supports that are not space-like separated, generally
$\left[\hat\phi_f,\hat\phi_g\right]\not=0$.
It is this nontrivial commutation relation that makes quantum optics a local quantum field theory.

If we now introduce a second set of creation and annihilation operators, $\hat b_g^\dagger$ and $\hat b_f$,
which commute with $\hat a_g^\dagger$ and $\hat a_f$ and which satisfy $\hat b_f\left|0\right>=0$ and
the commutation relations
\begin{eqnarray}
  \left[\hat b_f,\hat b_g^\dagger\right]&=&(g,f)_-
         =-\hbar\int k^\alpha\tilde g^*_{\alpha\beta}(k)\eta^{\beta\nu}
                                          k^\lambda\tilde f_{\lambda\nu}(k)
\makebox[3.5em]{$\theta(-k_0)$}\hspace{-3.5em}\rule[-1.25ex]{3.5em}{0.3ex}
                                          2\pi\delta(k_\mu k^\mu)\frac{\mathrm{d}^4k}{(2\pi)^4}\cr
         &=&(f^*,g^*),\cr
  \left[\hat b_f,\hat b_g\right]&=&0,
\end{eqnarray}
we can construct a random field observable 
\begin{equation}
  \hat\chi_f=\hat a_f+\hat a^\dagger_{f^*}+\hat b_f+\hat b^\dagger_{f^*},
\end{equation}
for which the commutator
\begin{eqnarray}
  \left[\hat\chi_f,\hat\chi_g\right]&=&\left[\hat a_f,\hat a^\dagger_{g^*}\right]
                                      +\left[\hat a^\dagger_{f^*},\hat a_g\right]
                                      +\left[\hat b_f,\hat b^\dagger_{g^*}\right]
                                      +\left[\hat b^\dagger_{f^*},\hat b_g\right]\cr
                                    &=&(g^*,f)-(f^*,g)+(f^*,g)-(g^*,f)=0
\end{eqnarray}
is trivial for all test functions $f$ and $g$; hence $\hat\chi_f$ can be called a classical random field.
The algebra of observables that is generated by $\hat\phi_f$ and $\hat\chi_f$ contains both quantum optics
and an associated classical random field optics.
Note that from a classical perspective negative frequency components are positive energy.

Many observables in quantum optics are not constructed as functions of $\hat\phi_f$, but are
instead constructed as projection operators such as $\hat a_f^\dagger\left|0\right>\left<0\right|\hat a_f$,
using only positive frequency test functions (with $f$ normalized, $(f,f)=1$).
We generally model both measurement and preparation apparatus as coupling to limited ranges of positive
frequency wave-numbers.
It makes no difference to any experimental predictions whether we use $a_f^\dagger$ or
$a_f^\dagger+b_f^\dagger$ to construct and measure physical states, if the test functions we use
are all purely positive frequency.
For all such observables, we can equally well discuss quantum optics or the associated classical random field
optics.

If all the local measurements we have so far made are, for systematic reasons, in the algebra generated by
$\hat\phi_f$, nonetheless quantum optics can be embedded in a larger field theory that has a classical
interpretation.
If we can in future construct devices that measure $\hat\chi_f$ as well as devices that measure $\hat\phi_f$,
the quantum theoretical models we construct could then be interpreted as a mathematics of classical
stochastic signal analysis\cite{CohenIEEE,CohenBook,Boashash}, in which quantum fluctuations of measurement
devices affect both the measured system and other measurement devices.
Even if we never succeed in measuring $\hat\chi_f$, we can nonetheless compute what the results of measurements
of $\hat\chi_f$ would be in the states that we in fact measure using only $\hat\phi_f$, and interpret quantum field
theory in terms of the enlarged algebra of observables.

In such an approach, Planck's constant is a measure of quantum fluctuations, which are classically distinguished
from thermal fluctuations by their Lorentz invariance properties\cite{MorganFluctuations}.
Planck's constant plays two r\^oles in a conventional free quantum field theory, both fixing the scale
of incompatibility $\left[\hat\phi_f,\hat\phi_g\right]=(g^*,f)-(f^*,g)$ between measurements and fixing
the scale of quantum fluctuations $\sqrt{(f^*,f)}$ of measurements of $\hat\phi_f$ in the vacuum state
(determined by the variance of the Gaussian characteristic function
$\left<0\right|e^{i\lambda\hat\phi_f}\left|0\right> = e^{-\lambda^2(f^*,f)/2}\;$).
These two r\^oles can be separated in two ways.
Firstly, we can introduce an observable
$\hat\xi_f=\alpha(\hat a_f+\hat a^\dagger_{f^*})+\beta(\hat b_f+\hat b^\dagger_{f^*})$, for which
\begin{eqnarray}
  \left[\hat\xi_f,\hat\xi_g\right]&=&(\alpha^2-\beta^2)\Bigl((g^*,f)-(f^*,g)\Bigr),\cr
  \left<0\right|e^{i\lambda\hat\xi_f}\left|0\right> &=& e^{-\lambda^2(\alpha^2+\beta^2)(f^*,f)/2}.
\end{eqnarray}
Secondly, we can introduce Lorentz invariant self-adjoint number operators $\hat\Xi_a$ and $\hat\Xi_b$,
which satisfy the commutation relations $\left[\hat\Xi_a,a^\dagger_f\right]=a^\dagger_f$ and
$\left[\hat\Xi_b,b^\dagger_f\right]=b^\dagger_f$, with all other commutation relations trivial.
The algebra generated by $\hat\Xi_a$, $\hat\Xi_b$, $\hat a_f^\dagger$, $\hat b_f^\dagger$,
$\hat a_f$ and $\hat b_f$ with these commutation relations satisfies the Jacobi identity.
$\hat\Xi_a$ and $\hat\Xi_b$ can be used, with the methods of \cite{MorganFluctuations,MorganWigner}, to
construct Lorentz invariant ``super''-vacuum states that have increased fluctuations, for which the
Gaussian characteristic function for the observable $\hat\xi_f$ is
\begin{equation}
  \frac{\mathrm{Tr}\left[e^{-\mu\hat\Xi_a-\nu\hat\Xi_b}e^{i\lambda\hat\xi_f}\right]}
       {\mathrm{Tr}\left[e^{-\mu\hat\Xi_a-\nu\hat\Xi_b}\right]}
    =e^{-\lambda^2(\alpha^2\coth{\mu}+\beta^2\coth{\nu})(f^*,f)/2},\qquad\mu,\nu>0.
\end{equation}
Both these constructions allow us to increase the scale of vacuum fluctuations relative to the scale
of measurement incompatibility, which we can tentatively characterize as
\begin{list}{$\bullet$}{\setlength{\itemsep}{0.5ex}\setlength{\topsep}{0ex}}
  \item suppressed fluctuations --- incompatibility between measurement devices is affected less by
        quantum fluctuations than the scale of quantum fluctuations of the vacuum, $\mu,\nu>0$;
  \item compensated fluctuations --- measurement devices interact with negative frequency quantum
        fluctuations, which compensate for the effects of positive frequency quantum fluctuations,
        $\beta^2>0$.
\end{list}
If we separate the two r\^oles of Planck's constant in either of these two ways, measurement
incompatibility becomes a property of measurement devices different from the scale of quantum
fluctuations.

For many years the stochastic electrodynamics approach (SED) has produced quite interesting results\cite{SED}
using a stochastic formalism for the electromagnetic field, however the present operator and Hilbert space
formalism for random fields is much closer to the familiar mathematics of quantum optics.
Both in SED and here, there are Lorentz invariant fluctuations of the electromagnetic field; merely by
introducing an explicit classical model for quantum fluctuations of the electromagnetic field, we go beyond
the capabilities of semiclassical optics.

When very many or infinite degrees of freedom are introduced, classical random field models for experiments
are of a very different nature from classical particle property models, so that the no-go theorems
against classical particle property models for quantum mechanical experiments are essentially irrelevant to
classical random field models.
I have previously shown that the violation of Bell inequalities, in particular, is not incompatible with
classical random field models\cite{MorganBell}, and contextuality is natural for classical random field
models --- by the inclusion of the experimental apparatus in models whenever necessary --- whereas contextuality
is not natural for classical particle property models.

Experiments in the quantum mechanical regime are often in coarse-grained equilibrium, in the sense that the
statistics of the discrete events that are measured are time-invariant.
Insofar as a given experiment is in coarse-grained equilibrium, a change of an experimental apparatus
is, to a first approximation at the level of statistics of discrete events, a change of the boundary
conditions of the electromagnetic random field at equilibrium, which generally has global consequences
for experimental results even if the dynamics is local.
Historically, Bohr and Heisenberg espoused a disturbance interpretation of quantum theory until Einstein,
Podolsky, and Rosen\cite{EPR} forced them, in 1935, to adopt an essentially positivistic
interpretation\cite[pp 34-35]{Fine}.
Particle properties cannot be given a local disturbance interpretation, but a local disturbance
interpretation of quantum field theory is possible if we take a random field approach.
Although measurements associated with space-time regions that are at time-like separation are apparently
incompatible with one another in general, as a matter of physical principle, we can mathematically model that
measurement incompatibility in a number of different ways.

There is an elementary sense in which the results presented above are not at all new.
The probability density associated with the observable $\hat\chi_f$ for any state is no more than the
possibly negative density associated with the observable $\hat\phi_f$ for the same state convolved with a
possibly negative density that is precisely enough to smear away the non-classical negative values,
which is a well-known but unmotivated and unappealing approach to reconciling the Wigner function, for
example, with classical preconceptions.
An algebraic context, however, gives some justification for and a way to interpret what is otherwise
a largely ad-hoc procedure.

A positive reason for using classical random field models in physics in the long term, once their
intrinsic properties, their measurement theory, and their relationship to quantum field theory are
better understood, is that a substantial class of interacting classical random fields can be constructed
straightforwardly\cite{MorganLieFields}, in contrast to the lack of any rigorous interacting
quantum field theory in four dimensions and the mathematical awkwardness of regularization and
renormalization.
It appears that measurement incompatibilities of a linear interacting field theory cannot be represented
by a local algebra of observables in which we consider only positive frequency components of test functions,
but we can use classical random field models of experiments that explicitly model interactions between
measurement devices that are due to quantum fluctuations.

\end{document}